# Multilayer Graphene Nanoribbon and Carbon Nanotube based Floating Gate Transistor for Nonvolatile Flash Memory

Nahid M. Hossain and Masud H Chowdhury

Computer Science and Electrical Engineering, University of Missouri – Kansas City, Kansas City, MO 64110, USA

Email: mnhtyd@mail.umkc.edu, masud@umkc.edu and masud@ieee.org

*Abstract-* Floating gate transistor is the basic building block of non-volatile flash memory, which is one of the most widely used memory gadgets in modern micro and nano electronic applications. Recently there has been a surge of interest to introduce a new generation of memory devices using graphene nanotechnology. In this paper we present a new floating gate transistor (FGT) design based on multilayer graphene nanoribbon (MLGNR) and carbon nanotube (CNT). In the proposed FGT a multilayer structure of graphene nanoribbon (GNR) would be used as the channel of the field effect transistor (FET) and a layer of CNTs would be used as the floating gate. We have performed an analysis of the charge accumulation mechanism in the floating gate and its dependence on the applied terminal voltages. Based on our analysis we have observed that proposed graphene based floating gate transistor could be operated at a reduced voltage compared to conventional silicon based floating gate devices. We have presented detail analysis of the operation and the programming and erasing processes of the proposed FGT, dependency of the programming and erasing current density on different parameters, impact of scaling the thicknesses of the control and tunneling oxides. These analysis are done based on the capacitance model of the device

## I. INTRODUCTION

Non-volatile flash memory that utilizes floating gate transistors (FGTs) has become the most widely used memory technology in numerous electronic applications. Due to continuous scaling and physical and material limits of conventional MOSFET technologies, silicon based floating gate transistors will no longer be able to meet the reliability, cost and efficiency requirements in future. Graphene that has extraordinary characteristics (very high carrier mobility, thermal conductivity, mechanical flexibility and strength, and optical transparency) is a highly promising material for future nonvolatile memory and other nanoelectronic devices [1][2]. The high carrier mobility of the MLGNR leads to the low latency and fast response. The intrinsic thermal conductivity protects the device from overheating. The mechanical flexibility inspires flexible memory, which is future of electronics design. In this paper we present the design of a new floating gate transistor using multilayer graphene nanoribbon (MLGNR) and carbon nanotube (CNT). The preliminary concept has been presented in our recent conference paper [26].

The primary difference between a floating gate transistor and the standard MOSFET is the addition of a new gate, called the floating gate, between the original gate and the channel as shown in Figure 1. The original gate (topmost) is now called the control gate. A floating gate is basically a polysilicon gate surrounded by insulator and it has no electrical connection with other layers [6]. The working principle of the floating gate transistor is almost same as conventional MOSFET, where the source-drain current is monitored and controlled by the control gate voltage. The floating gate voltage or in other words the stored charge on the floating gate can control the channel between the drain and the source. The thickness of the dielectric layer is around 10nm or less [8]. Thinner insulation layer is required to facilitate tunneling between the channel and the floating gate. The detail working principle of a floating gate transistor can be found in any relevant textbook. Interested reader can also refer to [6]-[8] and our recent conference paper [26] for further details.

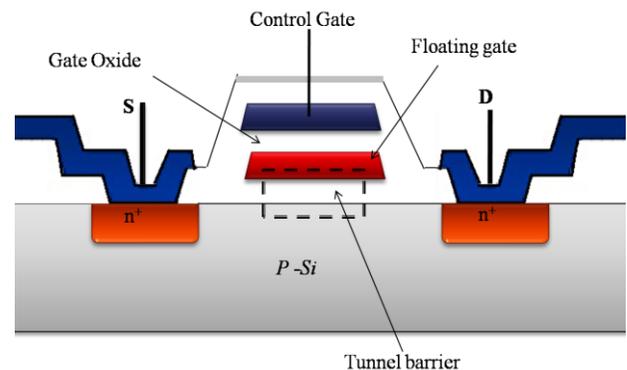

Figure 1: Schematic diagram of a floating gate transistor.

Many radical device and material alternatives are being explored for memory technologies in nanometer range. Nanoscale single-bit floating gate transistors and ZnO nanoparticle based floating gate transistor on very low cost glass and plastic substrate [11]-[13] for transparent electronics and memory devices are few examples. Floating gate transistor using gold nanoparticle and multiple-bits floating gate transistor have been reported in [3]. Graphene is another material that is getting widespread attention from diverse groups of engineers and scientists. The memory window of graphene based memory cell is expected to be greater than that of silicon [4]. Several graphene based memory cells have been under investigation. A floating gate transistor with few layers of graphene and molybdenum disulfide ($MoS_2$) as the channel material has been proposed in [9]. Graphene and molybdenum disulfide ($MoS_2$) were utilized as channel and charge trapping layers interchangeably, while hexagonal boron nitride was applied as tunnel barrier [10]. Graphene oxide thin films based flexible nonvolatile resistive memory has also been explored [5]. Graphene and graphene oxide have been investigated as FET channel, charge trapping layer and electrode in [14]-[19]. Large hysteresis in the gate characteristics of graphene FETs can be utilized for nonvolatile memory application [20]. The hysteresis arises from the oxide layer charge trapping [21].

In this paper, we present the concept of a new graphene based floating gate transistor, where the channel would be made of multilayer graphene nanoribbon (MLGNR) and the floating gate would be made of carbon nanotubes (CNT). The rest of the paper is organized as follows. Section II illustrates the design and construction of the proposed floating gate transistor based on MLGNR and CNT. This section also briefly highlights the potential fabrication process. Section III analyzes some of the basic physical and electrical parameters and its interdependence in the proposed devices. Section IV explains the programming and erasing processes of the proposed FGT. Section V investigates the impacts of scaling the insulating oxide layers around the gate. Section VI provides a comparison of the proposed graphene/CNT FGT with the existing FGT. Finally, Section VII concludes the paper with a brief overview of our ongoing work.

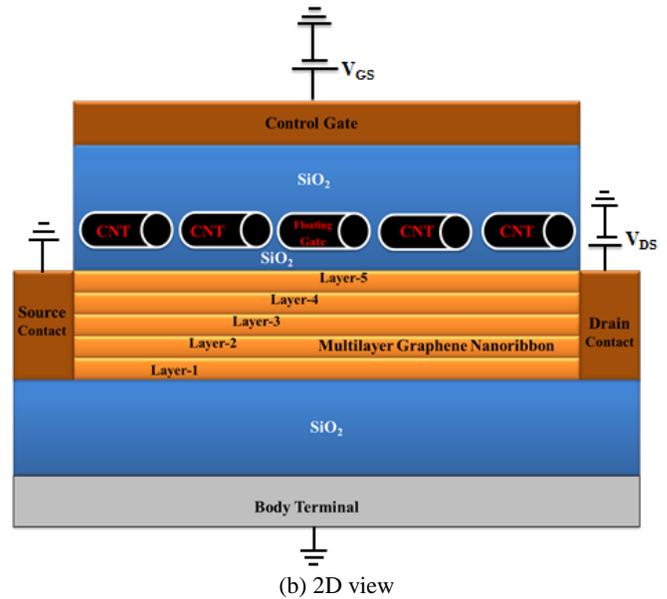

(b) 2D view

Figure 2: Proposed CNT and MLGNR based floating gate transistor with required electrical connection for its operation.

## II. PROPOSED GRAPHENE BASED FLOATING GATE TRANSISTOR

The proposed floating gate transistor is based on two forms of graphene (carbon) nanostructures, metal (polysilicon) contacts and $SiO_2$ insulator. Here MLGNR would be used as channel and CNTs would be used as electron trapping layer or floating gate. The schematic of our proposed graphene based floating gate transistor is shown in Figure 2. MLGNR is used in the proposed design to demonstrate the concept. Single layer graphene is not thermodynamically stable [42]. The single layer nano-patterned graphene FET is very noisy, while the nano-patterned few layer graphene FET shows reduced noise [43]-[44].

The potential fabrication process of the MLGNR/CNT FGT would include the following steps. First, a 300 nm thick $SiO_2$ layer is thermally grown on a silicon wafer, which is standard for graphene-based device. Second, MLGNR channel can be grown by Chemical Vapor Deposition (CVD) method, followed by an etching process to obtain a rectangular shape MLGNR with a uniform channel length and width. Third, $SiO_2$-CNT-$SiO_2$ sandwich are grown on the MLGNR sheet. Fourth, Ti/Au metal contact pairs can be used as the source, drain and control gate contact metal.

Although several layers of graphene is more attractive for obtaining low sheet resistance, beyond optimum number of GNR layers multi-layer graphenes convert into graphite [45]. As the number of GNR layers increase, effective resistance saturates which suggest that additional GNR layer are no longer improve resistance profile [46]. Therefore, the number of layers in the MLGNR structure would depend on the optimum performance requirements. Our future work will focus on the

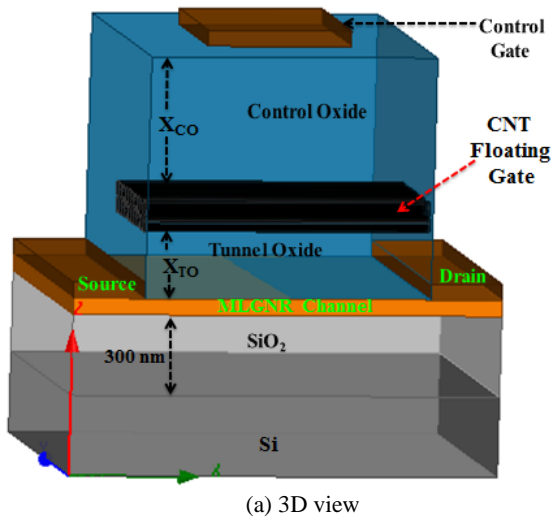

(a) 3D view

optimization of the proposed structure for the best possible performance. However, multiple GNRs would be required to provide strong conduction path and override noise. A layer of SiO$_2$ dielectric is placed between the MLGNR channel and body contact to provide electrical isolation. The body contact can also be configured as the second gate for better control.

Layers of SiO$_2$ are placed around the CNT layer to isolate the external control gate, the floating gate (CNTs) and the channel (MLGNR). The control oxide (the SiO$_2$ layer between the floating and the external gates) blocks the stored charge in the CNTs and prevents data loss resulting from charge leakage into the control gate electrode. Adoption of the control oxide effectively inhibits not only charge loss from the CNT charge-storage layer, but also blocks charge-injection from the metal control gate. This results in a higher trapping efficiency and relieves the problem caused by the thin charge storage layer. The tunnel oxide (the SiO$_2$ layer between the floating and the channel) must be thinner than the control oxide to allow electrons to smoothly tunnel to and from the channel and floating gate during the programming and erasing operation. Under normal operating condition the tunnel oxide has the same function as the control oxide to prevent charge flow in and out of the floating gate. Therefore, the dielectric and physical properties of the oxides around the gates are very critical for the performance and reliability of the proposed floating gate transistor. The external control gate (top gate) and the body contact would be made of metal, polysilicon or any other suitable conducting material. In future we will concentrate on other insulator like HfO$_2$ and contact materials.

The operation of this floating gate transistor would be similar to that of conventional MOSFET. The source and body terminals are connected to the ground and kept at the same potential to minimize leakage current. A drain-source voltage (V$_{DS}$) is applied to start the conduction in the MLGNR channel. Graphene channel offers three major advantages: (i) conduction starts at a very low voltage (~mV), (ii) undoped graphene can be used as channel material, and (iii) ambipolar conduction can be achieved. A positive gate voltage (V$_{GS}$) is applied to program the transistor. The value of V$_{GS}$ should be several times greater than V$_{DS}$. The specific value of V$_{GS}$ depends on the sense amplifier that determines the charge levels for '0' and '1' logic values. Equally distributed voltage is assumed across each GNR layer in the MLGNR channel. In this design we propose using undoped MLGNR. Usually, graphene is doped by atmospheric molecules, photoresist residue, metal etchants and Al$_2$O$_3$. Single layer graphene (SLG) is inherently p-type. On the other hand MLGNR is less sensitive to charge doping effects because the additional layers will lessen the effects of these charges [23]-[25]. Next section analyzes the electrical behaviors of the proposed FGT.

## III. BASIC PHYSICAL AND ELECTRICAL PARAMETERS OF THE PROPOSED DEVICE

In order to understand the dynamic behaviour of the proposed MLGNR/CNT floating gate transistor (FGT), its capacitive model has to be derived. Figure 3 shows the simplified capacitive model of the proposed FGT. Here, the C$_{FC}$, C$_{FG}$, C$_{FS}$ and C$_{FD}$ are the capacitances of the floating gate (FG) with the channel, control gate, source and drain. There will be three additional capacitances (C$_{SB}$, C$_{DB}$ and C$_{CB}$) inside the device. Here V$_{GS}$, V$_S$, V$_{DS}$, V$_C$ and V$_B$ are the potentials of the control gate, source, drain, channel and bulk respectively and V$_{FG}$ is the potential on the FG.

As the FG is connected with the control gate, source, drain and body terminals only through capacitors, the proposed flash memory cell can be expressed as a capacitor network as shown in Figure 3. According to (1), the total capacitance of the cell (C$_T$) is equal to the sum of the capacitances of the network. Here C$_{CB}$, C$_{SB}$ and C$_{DB}$ are not important in the model because the values of these capacitors are close to zero. These capcitances arise due to the substrate on which graphene is grown or transferred. The substrate is around 300nm thick that leads to negligible capacitance. The change of voltage (ΔV$_{FG}$) on the FG can be express as in (2). Here, Q$_{FG}$ is the total charge stored on the FG, which can be expressed by (3).

$$C_T = C_{FG} + C_{FS} + C_{FD} + C_{FC} \qquad (1)$$

$$\Delta V_{FG} = \frac{Q_{FG}}{C_{FG}} \qquad (2)$$

$$Q_{FG} = (V_{FG} - V_{GS})C_{FG} + (V_{FG} - V_S).C_{FS} \qquad (3)$$
$$+ (V_{FG} - V_D).C_{FD} + (V_{FG} - V_C).C_{FC}$$

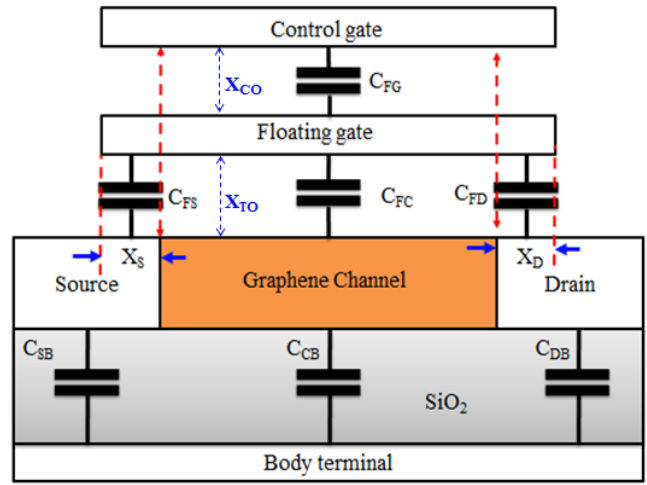

Figure 3: Capacitance model of the proposed MLGNR/CNT FGT considering fringing capacitance. X$_S$ and X$_D$ denote floating gate overlapping areas on source and drain respectively.

To derive the model for the voltage on the floating gate (V$_{FG}$) we can consider two cases.

*Case-1:* Floating gate with small overlaps with the source and drain regions. Consider the fringing capacitances (C$_{FS}$ and C$_{FD}$) between FG and the source and drain as shown in Figure 3. Using equations (1)-(3) we can derive the model for V$_{FG}$ as in (4).

$$Q_{FG} = V_{FG}(C_{FG} + C_{FS} + C_{FD} + C_{FC}) - V_{GS}.C_{FG}$$
$$- V_S.C_{FS} - V_D.C_{FD} - V_{Ch}.C_{FC}$$

$$Q_{FG} = V_{FG}(C_T) - V_{GS}.C_{FG} - V_S.C_{FS} - V_D.C_{FD}$$
$$- V_C.C_{FC}$$

$$V_{FG} . C_T = Q_{FG} + V_{GS}.C_{FG} + V_S.C_{FS} + V_D.C_{FD}$$
$$+ V_C.C_{FC}$$

$$V_{FG} = \frac{Q_{FG}}{C_T} + \left(\frac{C_{FG}}{C_T}\right).V_{GS} + \left(\frac{C_{FS}}{C_T}\right).V_S + \left(\frac{C_{FD}}{C_T}\right).V_D \quad (4)$$
$$+ \left(\frac{C_{FC}}{C_T}\right).V_C$$

*Case-2:* Floating gate is perfectly aligned with the graphene channel with no overlap between FG and source(drain) as in Figure 4. The fringing capacitances ($C_{FS}$ and $C_{FD}$) could be ignored. Then equation (3) can be simplified to (5). Again using equations (1) and (5) we can derive the model for $V_{FG}$ as in (6).

$$Q_{FG} = (V_{FG} - V_{GS})C_{FG} + (V_{FG} - V_C).C_{FC} \quad (5)$$

$$V_{FG} = \frac{Q_{FG}}{C_T} + \left(\frac{C_{FG}}{C_T}\right).V_{GS} + \left(\frac{C_{FC}}{C_T}\right).V_C \quad (6)$$

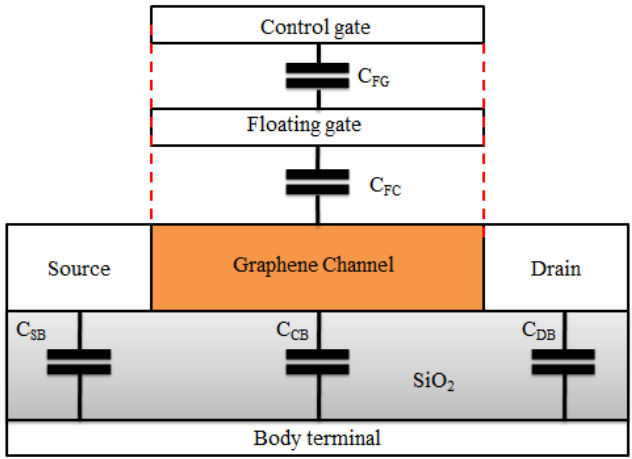

Figure 4: Capacitance model for FGT memory cell with no fringing capacitance between FG-drain and FG-source (Case-2).

For convenience, the coupling ratio terms, which are defined as the ratio of terminal voltage coupled to the floating gate, can be defined as follows:

GCR = control gate coupling ratio
DCR = drain coupling ratio
SCR = source coupling ratio
CCR = Channel coupling ratio

Thus, a variation in control gate voltage will result in a change in the floating gate voltage, $\Delta V_{FG} = \Delta V_{CG} \times$ GCR. The basic equation (4) for the capacitor network can be rewriiten in terms of the coupling ratio terms as in (7).

$$V_{FG} = \frac{Q_{FG}}{C_T} + (GCR.V_{GS}) + (SCR.V_S) + (DCR.V_D) \quad (7)$$
$$+ (CCR.V_C)$$

Initially, $Q_{FG}$=0, and for programming and erasing $V_S$=0V, $V_D \approx$ 0V. Therefore, $V_{FG}$=GCR.$V_{GS}$ because the term CCR.$V_C$ is negligible. The GCR is the key parameter, which defines the capacitive coupling ratio between the $C_{FG}$ and the $C_T$ as in (8). For programming GCR>0.60. The programming and erasing speed of flash memory depend on $V_{FG}$. So, it determines how fast a flash memory can be programmed and erased. The minimum programming and earse time of a device can be calculated from $V_{FG}$ value.

$$GCR = \frac{C_{FG}}{C_T} = \frac{C_{FG}}{C_{FG} + C_{FS}} \quad (8)$$

## IV. PROGRAMMING AND ERASING OPERATION OF THE PROPOSED FGT

In the proposed floating transistor the logic '0' and '1' states are determined by the programming and erase operations respectively. Under the influence of a positive control gate voltage electrons are accumulated on the floating gate (programming) that translates to logic state '0'. A negative voltage applied at the control gate leads to the depletion of electrons (erase) that translates to the logic state '1'. The electron accumulation and depletion are accomplished by tunneling - a process by which an electron passes through a barrier without physical conduction path. Ideally, an insulating oxide barrier doesn't allow charge to pass through it. However, at high electric field and thin oxide thickness tunneling takes place. The tunneling effect becomes more prominent as device dimensions enter deep into nanometer scale while electric field strength is on the rise as supply voltage scaling is slowed. While for non-memory device tunneling through gate oxide is an undesired phenomenon the operation of floating gate transistors in nonvolatile memory is dependent on tunneling. Therefore, analyzing the tunneling mechanism in the proposed MLGNR/CNT based FGT is a very critical part of understanding its programming and erasing operation and the evaluation of our concept.

The Fowler Nordheim (FN) mechanism is mostly used to realize programming and erasing current density ($J_{FN}$) of a floating gate transistor structure [29]. The programming and erasing tunneling current density ($J_{FN}$) can be calculated by (9)-(12). The parameters A and B depend on the work function or the barrier height ($\Phi_B$) at the interface between the tunneling oxide and the electron emitter and the effective mass of the tunneling electron $m_{ox}$. The work function is a property of the surface of the material. It depends on the crystal structure and the configurations of the atoms at the surface. A and B can be derived from FN plot ($J_{FN}/E^2$ vs. 1/E) as in [27]-[29]. Here, the induced electric field E is given by (10). By replacing E in (9) we get $J_{FN}$ as in (11). For source voltage $V_S$ =0V, $J_{FN}$ will be given by (12).

$$J_{FN} = AE^2 \exp\left[-\frac{B}{E}\right] \qquad (9)$$

Here, $A = \frac{q^3}{16\pi^2 h \varnothing_B}$ and $B = \frac{4}{3}\frac{(2m_{OX})^{\frac{1}{2}}}{qh}\varnothing_B^{\frac{3}{2}}$

$$E = \frac{V_{FG} - V_S}{X_{TO}} \qquad (10)$$

$$J_{FN} = A\left(\frac{V_{FG} - V_S}{X_{TO}}\right)^2 \exp\left[-\frac{B}{\left(\frac{V_{FG}-V_S}{X_{TO}}\right)}\right] \qquad (11)$$

$$J_{FN} = A\left(\frac{V_{FG}}{X_{TO}}\right)^2 \exp\left[-\frac{B}{\left(\frac{V_{FG}}{X_{TO}}\right)}\right] \qquad (12)$$

The subsequent paragraphs present the analysis of tunneling current during the programming and erasing operation of the proposed floating gate transistor based on the above models.

Figure 5 shows the dependence of the programming current density on the control gate voltage for a given control gate coupling ratio (GCR). This set of graph is generated from equation (7) and (12). As can be seen programming current increases with the increase of both the control gate voltage and GCR. Figure 6 shows the programming current variation with $V_{GS}$ for different tunnel oxide thickness ($X_{TO}$). It is observed that for a given $X_{TO}$, the programming current increases with $V_{GS}$. However, the programming current increases significantly when $X_{TO}$ is less than 7nm. According to ITRS 2011, semiconductor industry has already adopted 6nm tunneling oxide for 18-nm and 22-nm technology nodes. While 5nm tunnel oxide is predicted for 8-14nm technology nodes. Therefore, for technology nodes below 20nm, high programming current density will affect the reliability of the tunnel oxide.

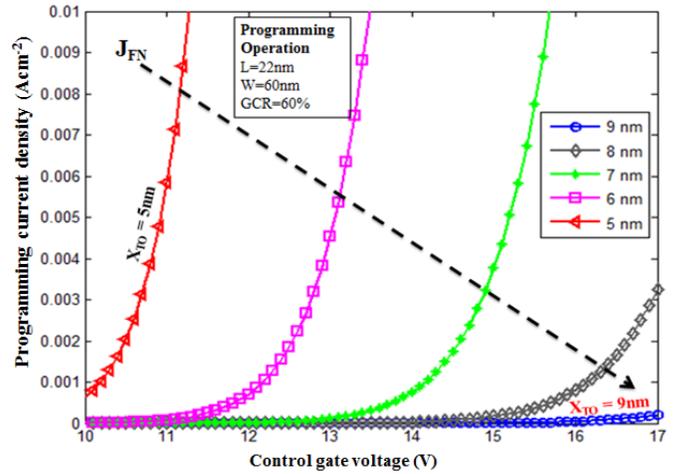

Figure 6: The programming current density versus control gate voltage for five different tunnel oxide thicknesses ($X_{TO}$). Here, GCR=60%, $V_{GS}$ =10-17V.

During the erasing operation a negative voltage would be applied at the control gate. We have performed the same set of analysis (as in Figure 5 and Figure 6) for the erasing operation. Figure 7 shows that erasing current increases as the control gate voltage ($V_{GS}$) becomes more negative for a given GCR. Higher GCR leads to higher current density because large control gate coupling will increase electron depletion rate from the floating gate to the MLGNR channel. Figure 8 shows the erasing current variation with $V_{GS}$ for different $X_{TO}$. It is seen erasing current density increases with the increase of $V_{GS}$ in the negative direction for a given $X_{TO}$. The tunneling current increases significantly when $X_{TO}$ is less than 7nm similar to the programing operation.

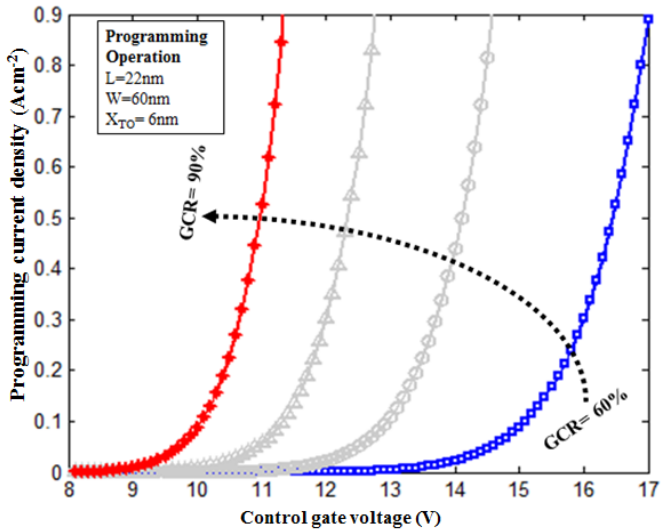

Figure 5: The programming current density versus control gate voltage for four different GCR.

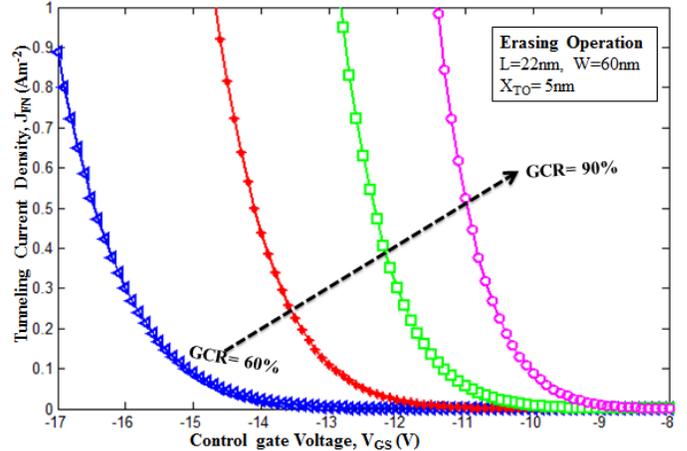

Figure 7: The erasing current density versus Control gate voltage for four different GCR (%). $X_{TO}$=5, $V_{GS}$ <0V.

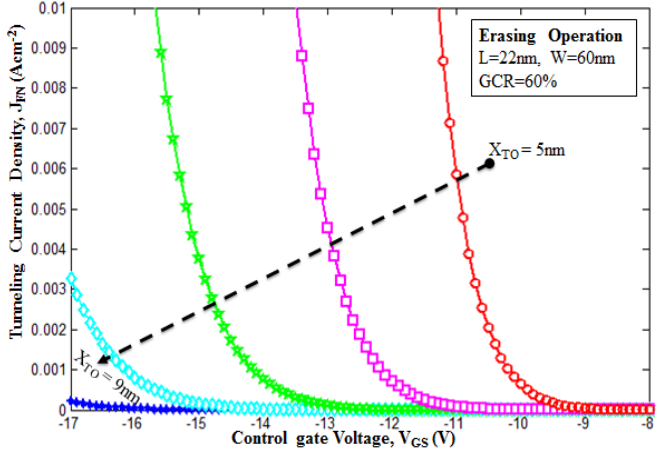

Figure 8: The erasing current density versus control gate voltage for five different tunnel oxide thicknesses ($X_{TO}$) GCR=60%, $V_{GS}$ <0V.

## V. ANALYSIS OF OXIDE THICKNESS SCALING IN MLGNR/CNT FLOATING GATE TRANSISTOR

The programming and erasing speed of flash memory depend on the voltage at the floating gate ($V_{FG}$), which determines how fast a flash memory can be programmed and erased. The minimum programming and earse time of a device can be calculated from $V_{FG}$ value. The control gate coupling ratio (GCR) controls the $V_{FG}$ according to the equation (7). Conventionally, GCR measures the percentage of the control gate voltage that is coupled from the control gate to the floating gate. For faster programming higher GCR is expected. For conventional silicon based FGTs GCR>0.6 is standard. For our proposed MLGNR/CNT FGT we are anticipating similar range. However, as we focus on the optimization of the device geometry and material composition in the proposed design we will address the recommended range for GCR.

In the next two subsections the impacts of the scaling of control oxide and tunnel oxide thicknesses are analyzed.

### A. Impacts of Scaling the Thickness of Control and Tunnel Oxides on GCR

The capacitance between the control gate and the floating gate is given (13). Here, A is the effective area of the floating gate and $\varepsilon_{CO}$ is the permittivity of the control oxide.

$$C_{FG} = \frac{A. \varepsilon_{CO}}{X_{CO}} \qquad (13)$$

The capacitance between channel and floating gate is

$$C_{FC} = \frac{A. \varepsilon_{TO}}{X_{TO}} \qquad (14)$$

Here, A is the effective area of the floating gate and $\varepsilon_{TO}$ is the permittivity of the tunnel oxide. Replacing $C_{FG}$ in (3) by (4) the model for GCR can be found as in (6).

$$GCR = \frac{\frac{A. \varepsilon_{CO}}{X_{CO}}}{C_{FG} + C_{FS} + C_{FD} + C_{FC}} \qquad (15)$$

The Variation of GCR with the variation of the thickness of the control oxide ($X_{CO}$) is plotted in Figure 9 for four different values (5nm, 6nm, 8nm and 10nm) of tunnel oxide thickness ($X_{TO}$). It is observed from Figure 9 that with the scaling down of the control oxide GCR value increases exponentially for a given tunnel oxide thickness ($X_{TO}$). This suggests that GCR in the proposed MLGNR/CNT FGT is a strong function of $X_{CO}$. Therefore, if the control gate requires more control than other terminals on the device operation, the control oxide should be scaled down. But the control oxide should not be scaled down less than two times of the tunnel oxide thickness ($X_{CO} > 2X_{TO}$) because this terminal is very prone to leakage, noise and radiation. Figure 10 shows the variation of GCR with the scaling of tunnel oxide thickness ($X_{TO}$) for different values (12nm, 20nm and 30nm) of $X_{CO}$. It is observed that the value of GCR decreases with the scaling down of $X_{TO}$ for a particular value of $X_{CO}$.

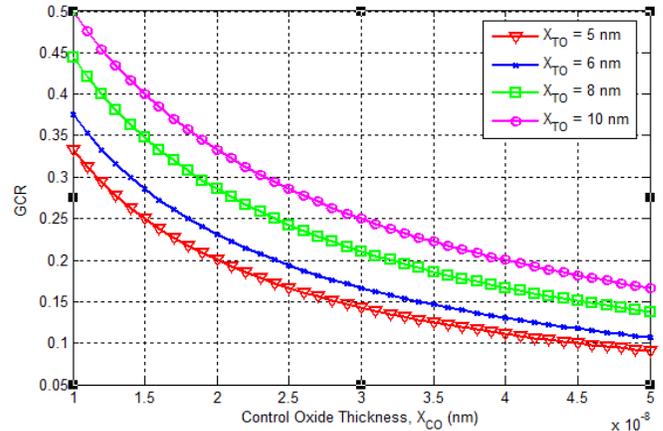

Figure 9: GCR versus $X_{CO}$ for different $X_{TO}$.

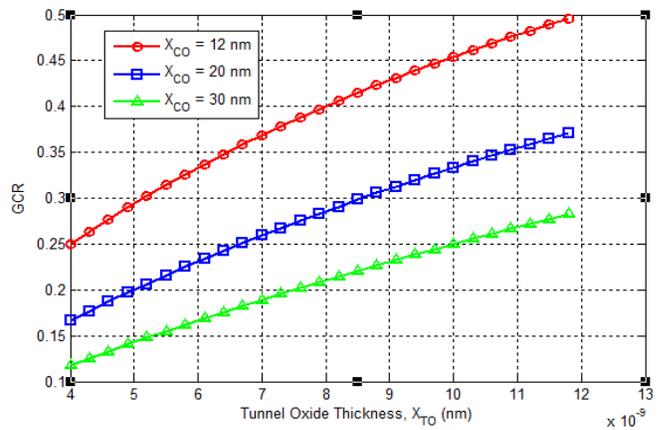

Figure 10: GCR versus $X_{TO}$ for different $X_{CO}$.

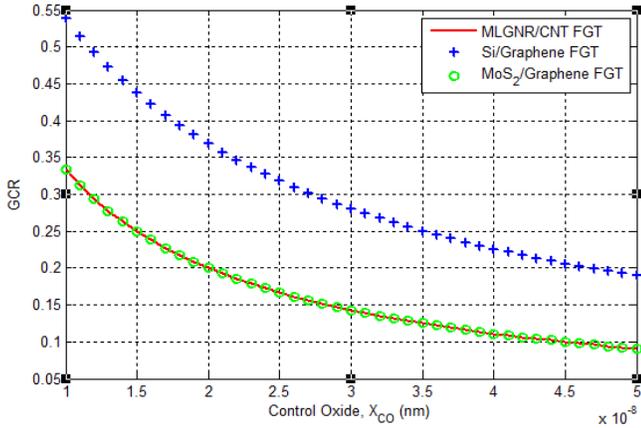

Figure 11: GCR-$X_{CO}$ analysis comparison of the MLGNR/CNT FGT with the existing graphene based FGT devices.

Finally, GCR-$X_{CO}$ and GCR-$X_{TO}$ analysis comparisons of the MLGNR/CNT FGT, with the existing graphene based FGTs, are shown in Figure 11 and Figure 12. Although high k-dielectric HfO$_2$ control oxide and tunnel oxide have been used in MoS$_2$/Graphene FGT [40] design, it exhibits similar GCR variation of the MLGNR/CNT FGT. On the other hand Si/Graphene [39] FGT shows better performance than the other designs due to its high k-dielectric control oxide (Al$_2$O$_3$) and low k-dielectric tunnel oxide (SiO$_2$). Therefore, it can be concluded that high k-dielectric control and low k-dielectric tunnel oxide can improve the GCR value.

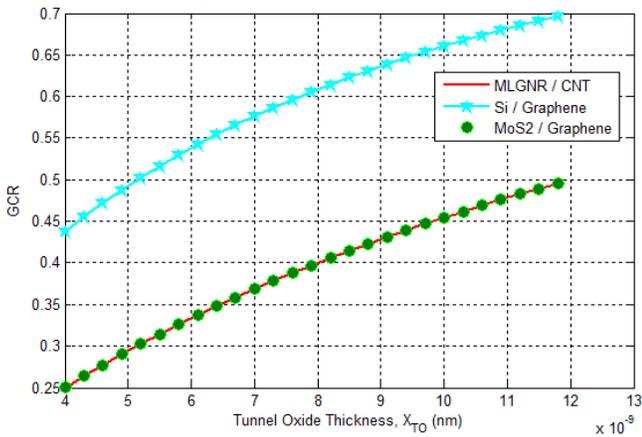

Figure 12: GCR-$X_{TO}$ analysis comparison of the MLGNR/CNT FGT with the existing graphene based FGT devices.

### B. Impacts of Scaling the Thickness of Control and Tunnel Oxides on CCR

The Channel coupling ratio (CCR) is defined as the ratio of the channel voltage coupled to the floating gate. This is measured by the ratio of the capacitance between the floating gate and channel (C$_{FC}$) to the total capacitance of the FGT (C$_T$) as in (16).

$$CCR = \frac{C_{FC}}{C_T}$$

$$CCR = \frac{\frac{A.\varepsilon_{TO}}{x_{TO}}}{C_{FG} + C_{FS} + C_{FD} + C_{FC}} \quad (16)$$

The variation of the CCR-$X_{TO}$ is plotted in Figure 13. According to the graph, as the tunnel oxide scales down, the CCR increases for a given control oxide thickness ($X_{CO}$=12nm, 20nm, 30nm) (Figure 13). The variation of the CCR-$X_{CO}$ is illustrated in Figure 14. It is observed that the CCR decreases for a given tunnel oxide thickness ($X_{TO}$= 5nm, 6nm, 8 nm, 10nm) when the control oxide is scaled down.

It is clearly shown that the GCR and CCR of the MLGNR/CNT FGT are strong function of the control oxide and tunnel oxide. Therefore, if the control gate (channel) needs more control than other terminals on the device, control oxide should be scaled down. So, the GCR decreases when the CCR increases, but the GCR increases when the CCR decreases. The relationship between GCR and CCR can be determined by the simple formula, GCR+CCR+SCR+DCR=1. But the tunnel oxide should not be scaled down less than 6nm, which will increase the tunneling current significantly. The 6nm tunnel oxide thickness (SiO$_2$) is very popular for floating gate transistor in current semiconductor industry.

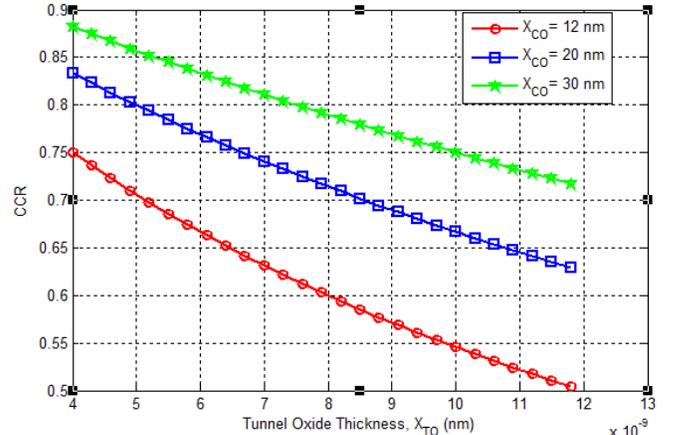

Figure 13: CCR versus $X_{TO}$ for different control oxide thicknesses.

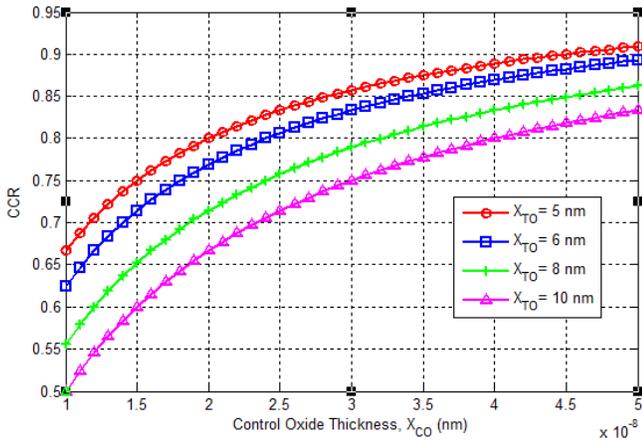

Figure 14: CCR versus $X_{CO}$ for different tunnel oxide thicknesses.

## VI. COMPARISON WITH THE EXISTING FGTs

To justify the prospects of the new floating gate transistor (FGT) design we have performed an analysis of the the relative performances of the emerging FGT devices (Si/Graphene, MoS₂/Graphene and MLGNR/CNT) and the conventional silicon FGT under identical conditions. Initially, there are no charges on the FG, ($Q_{FG}=0C$) and the body terminal is connected to the ground ($V_B=0V$). Therefore, $V_{FG}$ depends on GCR and $V_{GS}$. Our analysis shows that the MLGNR/CNT FGT couples more $V_{FG}$ than the conventional Si-FGT (Figure 15).

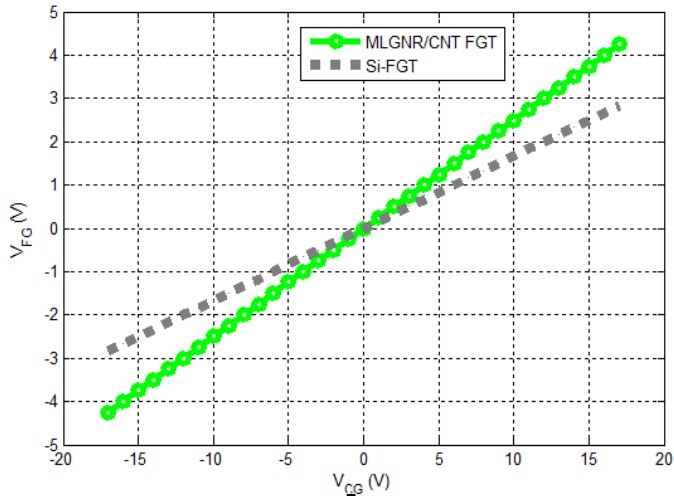

Figure 15: The $V_{FG}$ comparison between the MLGNR/CNT FGT [26] and the conventional Si-FGT [34]. The result shows good agreement with the existing experimental works [34], [26].

In order to compare the Si/Graphene, MoS₂/Graphene, MLGNR/CNT and conventional silicon FGTs, the same program-erase voltage (±17V) is applied. The simulation result shows that our proposed MLGNR/CNT FGT outperforms other flash memory designs (Figure 16). Our analysis shows that our proposed MLGNR/CNT FGT is able to couple more $V_{FG}$ than the MoS₂/Graphene and the Si-FGT. In the proposed design the thickness of the control oxide can be reduced and optimized to achieve high value of $C_{FG}$. Therefore, high GCR can be achieved, which leads to further increse of $V_{FG}$. Again, the performance of our MLGNR-CNT FGT can be boosted, by using high k dielectric material (HfO₂/Al₂O₃) as the control oxide while low k dielectric material (SiO₂) should be used as the tunnel oxide.

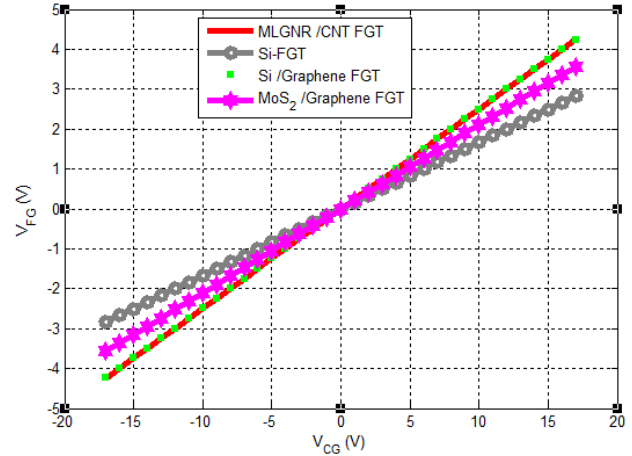

Figure 16: The $V_{FG}$ comparisons of the proposed MLGNR/CNT [26] FGT with the IBM Si/Graphene [39], MoS₂/Graphene[40], conventional Si FGT [34] flash memories. Our Proposed MLGNR/CNT FGT shows better performance than other existing designs. The result shows good agreement with the existing experimental works [40],[34]-[39].

## VII. CONCLUSION AND FUTURE WORK

The proposed floating gate transistor for nonvolatile memory has the potential to utilize all the excellent electrical, physical, thermal and material properties graphene nanoribbon and carbon nanotube. In general graphene nanotechnology is getting widespread attention for the next generation logic and memory devices for various nanoelectronic applications. The proposed MLGNR-CNT FGT is expected to open the door for a class of memory devices in this sector. Our preliminary concept was presented in crude form in our upcoming conference paper [26]. Here we provided detail description of the design and the underlying scientific explanation behind the concept. We have performed analysis of the electrical behaviors and dynamic characteristics of the device. We have also derived the capacitive model of the device. Through our analysis we have identified the critical electrical, physical and geometrical parameters that would impact the operation and performance of the device.

It is concluded for faster programming and erasing higher FN tunneling current density ($J_{FN}$) can be achieved by higher control gate voltage and scaling down the thicknesses of the control gate oxide and tunnel oxide. However, higher tunneling current will severely damage the oxide's reliability. Therefore, an optimization among these crucial parameters is recommended. Our future work will involve optimizing the supply voltage, tunneling current density and oxide thickness for optimum performance. Also, more accurate models for $J_{FN}$ and other electrical behaviors need to be developed.

The scaling of the control and tunnel oxides in the proposed MLGNR/CNT FGT is discussed in details. It is clear that the coupling capabilities of the control gate and the channel are the functions of both the control and the tunnel oxides' thicknesses. In other word, if the tunnel oxide is scaled down, the CCR rises while the GCR drops. On the other hand, if the control oxide is scaled down the GCR increases while the CCR falls. From these discussions, the 6nm tunnel oxide thickness and greater than 12nm control oxide thickness are recommended for the MLGNR/CNT FGT. The above statement is again supported by the tunneling current analysis through the tunnel oxide, which states that the tunneling current increases significantly when tunnel oxide ($SiO_2$) is scaled down from 6nm to 5nm. Finally, the GCR of the MLGNR/CNT FGT can be further improved by using high-k dielectric oxide on the control side and low-k dielectric oxide on the tunnel side of the gate. This would obviously increase design complexity. Our future work will focus on alternative to $SiO_2$ as control oxide.

The proposed concept of a new MLGNR-CNT based FGT seems promising. However, the investigation is still far from the final stage. There are many issues and challenges that need to be analyzed and resolved before validating the design through a set of real experimentations. The whole simulation and analysis process has been conducted using MATLAB. We have neglected three issues to simplify our simulation. First, we considered no gap between two adjacent graphene layers. It is still now unresolved whether there has to be any gap or separating material between two adjacent GNRs in a MLGNR structure. Although there are many MLGNR and CNT based transistors currently under investigation for both logic and memory devices, researchers are still investigating various aspects of any MLGNR and CNT based design. Second, same voltage level is considered for all layers in the MLGNR structure. Third, contact resistance of graphene-Cu interface is ignored. We are currently conducting a separate research project to model and characterize graphene-copper contact properties and resistance. It is also important to investigate better contact materials other than copper. If successful in finding a suitable contact material and mechanism for CNT and GNR this alone would be a groundbreaking work for all current and future work on graphene based nanoelectronic devices.

Our analysis reveals that the proposed device is capable to accumulate minimum required charge at very low voltage, which is a direct indication of low power operation. It is observed that the control gate voltage is solely responsible for tunneling and accumulating electron in the floating gate. Many contemporary works indicate that graphene as floating gate material has good charge retention capacity. However, a separate and thorough investigation has to be conducted on the charge retention capacity and optimization techniques for CNTs and GNRs in any memory device including the proposed one in this paper. Our ongoing research includes detail analysis of the charge retention capability, current-voltage (I-V) characteristics, doping requirements (if any) and manufacturing and implementation approaches for the proposed graphene based FGT.